\documentclass[a4paper]{llncs}
\usepackage{graphicx}
\usepackage{amsmath}
\usepackage{hyperref}
\usepackage{color}
\usepackage{subfigure}

\begin{document}

\title{How the competitive altruism leads to bistable homogeneous states of cooperation or defection}

\author{A. Jarynowski$^1$, P. Gawro\'nski$^2$ and K. Ku{\l}akowski$^2$}

\institute{$^1$ Faculty of Physics, Astronomy and Applied Computer Science, Jagiellonian University, ul. Reymonta 4, PL-30059 Krak\'ow, Poland\\ $^2$Faculty of Physics and Applied Computer Science, AGH University of Science and Technology, al. Mickiewicza 30, PL-30059 Krak\'ow, Poland \\
\email{andrzej.jarynowski@uj.edu.pl, gawron@newton.ftj.agh.edu.pl, kulakowski@novell.ftj.agh.edu.pl}
}

\maketitle

\begin{abstract}
Our recent minimal model of cooperation (P. Gawronski et al, Physica A 388 (2009) 3581) is modified as to allow for time-dependent altruism. This evolution is based on reputation of other agents, which in turn depends on history. We show that this modification leads to two absorbing states of the whole system, where the cooperation flourishes in one state and is absent in another one. The effect is compared with the results obtained with the model of indirect reciprocity, where the altruism of agents is constant.
\end{abstract}


\section{Introduction}

The Prisoner's Dilemma (PD) is a canonical example of game, where mutual cooperation is not profitable for an individual player and simultaneously it is 
profitable for a society. This paradoxical aspect and the wide set of situations where PD applies makes it a central point in game theory \cite{gin}.
On the sociological side, PD is considered as one of three generic games which represent three paradigmatic interaction situations; coordination and inequality are two others \cite{eum}. As a rule, PD is investigated in the frames of game theory, i.e. decisions are made on the basis of calculated payoffs. However, many scholars indicated that these frames are too narrow \cite{eum,cbb,guk,eo2}. In particular in \cite{fegi}, an opposition of two archetypal players has been presented: Homo Economicus - a creature who is rational and purely self-regarding, and Homo Sociologicus - a creature who follows prevailing social norms, and both concepts have been stated to be erroneous as one-sided. The truth, if accessible, is supposed to be in between. Still, any receipt how to weight these two attitudes in a given society remains arbitrary. It seems obvious that any classification of this kind is improved when we accept a gradual scale.\\

Here we are interested in a model of possible cooperation where both aspects are captured, the normative one and the rational one. Our starting point is the social mechanism of competitive altruism \cite{rob,rob2}, when individuals compete for most altruistic partners; in this competition, altruistic behaviour is a signal. This mechanism was the basis of our previous minimal model of cooperation, where players differed in altruism; the latter was defined as a willingness to cooperate \cite{gaw}. The agents' behaviour was encoded in the form of their time-dependent reputation. Our nice result was that altruistic players were rewarded by cooperation of other agents. In this way, the classical deficiency of altruistic strategy - being abused by selfish free-riders - was evaded in our model. \\

Simple as it was formulated, the model of cooperation \cite{gaw} is indeed minimal in the sense that its outcome is obtained with a minimal number of assumptions. The aim of this paper is to built into the model the fact that people learn, and their willingness to cooperate is modified by their personal experience. The dynamics of one's willingness to cooperate is at the core of the phenomenon of competitive altruism. In \cite{rob2}, we read:
"Competitive altruism is based on two simple premises. First it assumes that there are individual differences in altruism. (...) Second, in forming alliances there is competition for the most moral and cooperative partners." In our previous formulation \cite{gaw}, only the first premise was included directly: the players differed in the values of their altruism. However, it is clear that in competing with others, one must modify own behaviour as to outperform the rivals. Therefore, individual altruisms should vary as well, if the second premise is to be included to the model.\\

With this premise in mind, we propose here two schemes of varying the players' altruism after each game. According to our first option, say A, altruism is updated in the same way as the reputation in \cite{gaw}, only somewhat slower. This small difference seems realistic: we change our opinions on other people almost immediately after getting new information about them. Also, the difference is dictated by our aim to compare the results with those in \cite{gaw}, where altruism of each agent was constant. Second option B introduces a modification which is suggested by Scheff theory of shame and pride \cite{sch}. According to this theory, a mutual respect of two agents expressed by their cooperation enhances their self-evaluation, what in turn reinforces their willingness to cooperate. On the other hand, a cooperating agent is humiliated when mets a defection, what reduces her/his willingness to cooperate.\\

How these modifications of altruism influence the ability of the population to cooperate? In the option A, strategies of cooperation and of defection are equivalent, then the solution should be symmetric with respect to interchange of these strategies. As we explain in detail in the next section, in the option B two succesful cooperators have their altruisms increased, but this event can happen with the probability equal to, roughly, squared concentration of cooperators. If this concentration is initially 1/2, the process should be neutralized by a decrease of altruism of an unsuccessful cooperator. Below we show that this is not the case; the rules, apparently symmetric, promote the cooperation.\\
 
Actually, it seems that the coupling between the willingnesses to cooperate of different players is more general than the effect of competitive altruism or of the loops of shame and pride. This coupling can be viewed as a general ability of a set of players to establish a social norm of cooperation, and this norm in turns allows to define a social group. The role of norms in establishing social groups and societies is much too wide a subject to be discussed here \cite{guk,deb}. Instead, we refer only to the definition of a social norm \cite{cb}, which clearly underlines the role of mutual expectation of agents: once they believe they recognize the attitudes of the others, a germ of a norm is established. \\

This ambiguity suggests, that it is desirable to look for another mechanism, not due to the modification of altruism, which leads to an enhancement of cooperation. Our choice is to check the scenario where altruism of each agent is constant, but the modification of reputation depends on the reputation of co-player. In this way, the parameter controlling the change of reputation of agent $i$ is just the reputation of $i$'s coplayer $j$. This choice, marked here as option C, is motivated by the mechanism of indirect reciprocity and punishment, discussed e.g. in \cite{oht}. In other words, the modification of reputation in a game with an agent with bad reputation remains small.\\

The outline of the paper is as follows. In the next section we explain the original version of the model \cite{gaw} and options A, B and C considered here. In the third section numerical results are described. There we demonstrate that when altruism is varied, all players defect or all cooperate in the stationary state. This bistability with two homogeneous states in the main result of the paper. We show also, that for the option C, cooperation is promoted. The last section is devoted to discussion. 

\begin{figure}[ht]
 \centering
 {\centering \resizebox*{12cm}{9cm}{\rotatebox{-90}{\includegraphics{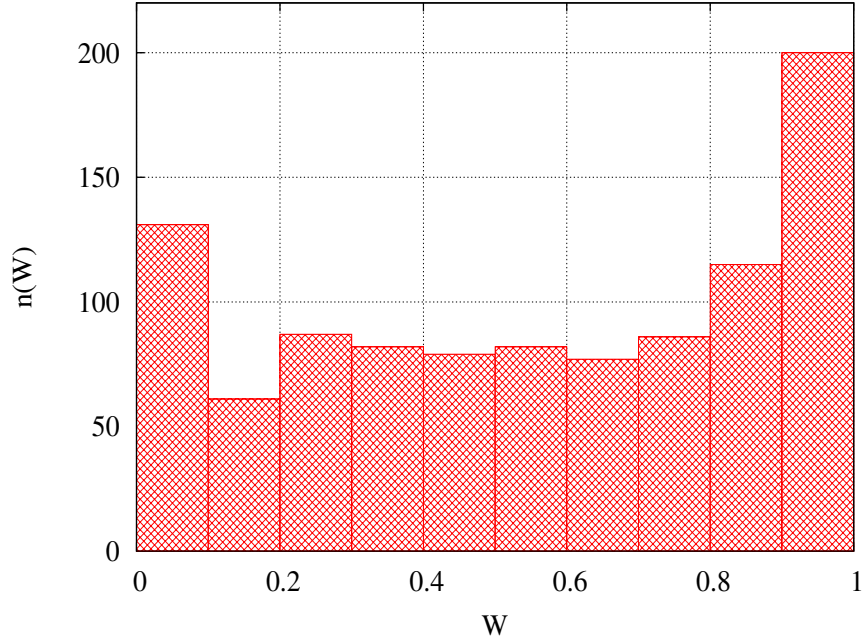}}}}
\caption{The reputation distribution $n(W)$ for an exemplary system of $N=10^3$ agents, evolving according to the option A, at the early stage after $10^4$ games. Here, the initial distribution of altruism is symmetric (case {\it ii)}). After a relatively quick development of groups of agents with good and bad reputation, slowly those of good reputation start to dominate - the right maximum grows larger.}
 \label{fig-1}
 \end{figure}

\begin{figure}[ht]
 \centering
 {\centering \resizebox*{12cm}{9cm}{\rotatebox{-90}{\includegraphics{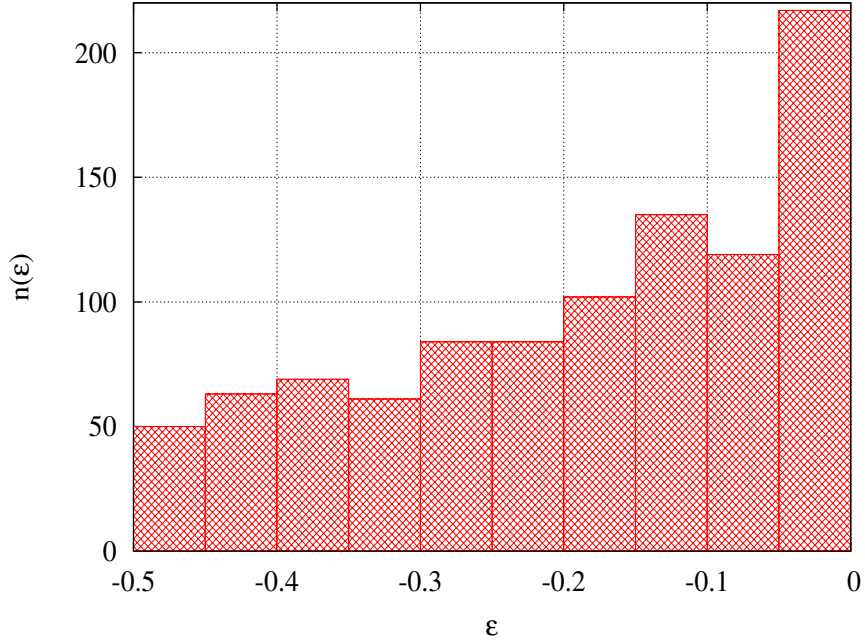}}}}
\caption{The altruism distribution $n(\varepsilon)$ for an exemplary system of $N=10^3$ agents, evolving according to the option B. After $10^5$ games, the distribution is almost stable. Here, the initial distribution of altruism promotes defection (case {\it i)}).}
 \label{fig-2}
 \end{figure}

\begin{figure}[ht]
 \centering
 {\centering \resizebox*{12cm}{9cm}{\rotatebox{-90}{\includegraphics{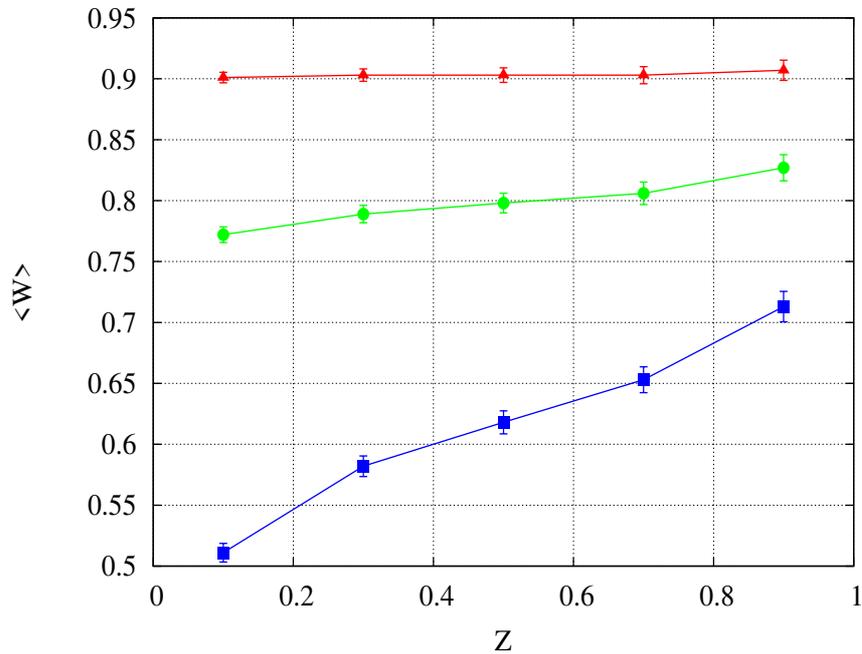}}}}
\caption{Mean value of reputation after $10^5$ games for a system of $N=10^3$ agents evolving according to the option C, against the parameter $z$ which controls the speed of evolution of reputation. Three curves are for three cases of initial conditions, $\it i), ii)$ and {\it iii)} from the bottom to the top. The statistics is collected for $10^3$ systems.}
 \label{fig-3}
 \end{figure}

\begin{figure}[ht]
 \centering
 {\centering \resizebox*{12cm}{9cm}{\rotatebox{-90}{\includegraphics{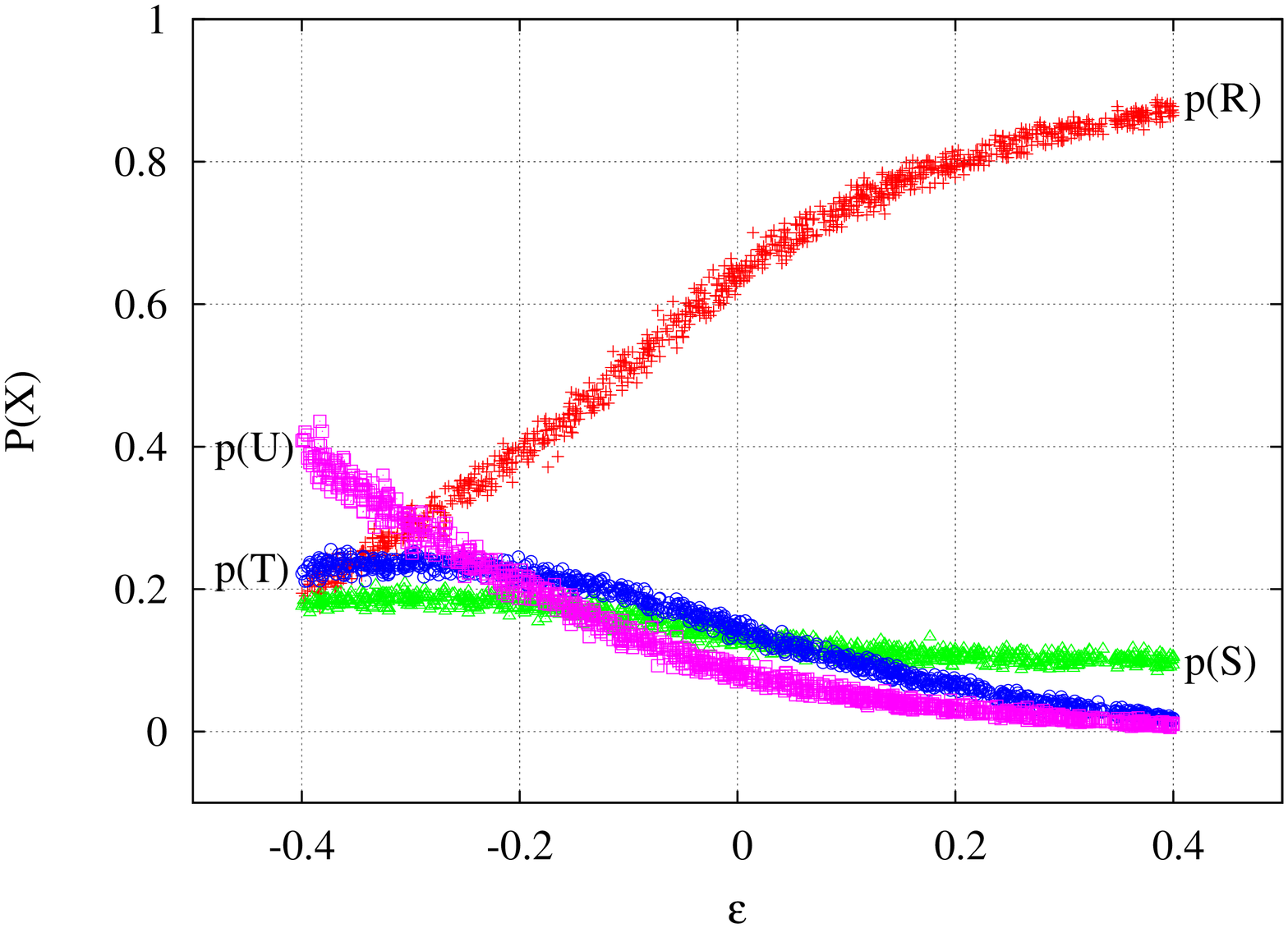}}}}
 \caption{The probabilities of four outcomes of Prisoner's Dilemma game for the system evolving according to the option C.  Here, the initial distribution of altruism is symmetric (case {\it ii)}). The statistics is collected for an exemplary system of $N=10^3$ agents after $10^6$ games. }
 \label{fig-4}
 \end{figure}

\section{The model}

The system is equivalent to a fully connected graph of $N$ nodes, with a link between each pair of agents at the nodes. Each agent $i$ is endowed with two parameters: altruism $\varepsilon_i$ and reputation $W_i$.  Initial values of the parameters are selected randomly from homogeneous distributions: $\rho(\varepsilon_i)$ is constant for $E_1<\varepsilon_i<E_2$, otherwise $\rho(\varepsilon_i)=0$, and $\rho(W_i)$ is constant for $V_1<W_i<V_2$, otherwise $\rho(W_i)=0$. \\

During the simulation, a pair of nodes $(i,j)$ is selected randomly. The probability that $i$ cooperates with $j$ is 

\begin{equation}
P(i,j)=F(\varepsilon_i+W_j)
\end{equation}
where $F(x)=0$ if $x<0$,  $F(x)=x$ if $0<x<1$ and $F(x)=1$ if $x>1$. In \cite{gaw}, it was only reputations $W_i$, $W_j$ what evolved in time. If $i$ cooperated, her/his reputation was transformed as $W_i \to (1+W_i)/2$, otherwise $W_i \to W_i/2$. Here we set $W_i$ to change in the same way.\\

Moreover - and this is a new element - we allow also the altruism to change. In the option A, this change is ruled according to a similar prescription. If $i,j$ play, the altruism of $j$ varies as

\begin{equation}
\varepsilon_j \to \varepsilon_j+(\pm1/2-\varepsilon_j)x
\end{equation}
where $0<x<1$ is a parameter which measures the velocity of change of altruism, the sign $+1$ applies if $i$ cooperates and the sign $-1$ - if $i$ defects. In other words, the altruism $\varepsilon_j$ increases if $j$ mets cooperation, decreases otherwise. As long as $x<1/2$, the time evolution of altruism is slower than the one of reputation. Here we set $x=0.1.$ Larger values of $x$ just speed up the changes of agents' altruism.\\

In the option B, it is only the rule of variation of $\varepsilon_i$ what is changed with respect of the option A. Namely, if both $i$ and $j$ cooperate, their altruism increase as

\begin{eqnarray}
\varepsilon_i \to \varepsilon_i+(1/2-\varepsilon_i)x\\
\varepsilon_j \to \varepsilon_j+(1/2-\varepsilon_j)x
\end{eqnarray}
If $i$ cooperates and $j$ defects, then the altruism of $i$ is reduced as

\begin{equation}
\varepsilon_i \to \varepsilon_i+(-1/2-\varepsilon_i)x
\end{equation}
whilst $\varepsilon_j$ is not changed. When both $i$ and $j$ defect, nothing is changed.\\

In the option C, altruism remains constant, as in \cite{gaw}. The velocity of the variation of reputation of $i$ is controlled by the reputation of her/his coplayer $j$. Namely, when $i$ cooperates, then her/his reputation $W_i$ changes as

\begin{equation}
W_i \to W_i(1-zW_j)+zW_j
\end{equation}
where $z$ is a parameter. When $i$ defects, 

\begin{equation}
W_i \to W_i(1-zW_j).
\end{equation}

\section{Results}

For the option A, the problem is symmetric with respect to an interchange of the strategies: cooperation and defection. When also the initial distributions of both reputation and altruism are symmetric ($V_1+V_2=1$ and $E_1+E_2=0$), this symmetry should be preserved also in the solution. This is so, however, only in the statistical sense. For each simulation, the system breaks the symmetry spontaneously and the time evolution leads to one of two homogeneous states, where each agent adopts the same strategy. In one of these two states, all agents cooperate; in another, all defect. The process of spontaneous symmetry breaking is visible in Fig. 1. There, we show the probability distributions of reputation and altruism for $N=10^3$ agents at the early stage of the process, after $10^4$ games. For the symmetric initial state where $\bar\varepsilon=0$ and $V_1=V_2=0.5$, the result obtained numerically as an average over $10^3$ systems, each after $10^5$ games, is that the probability of the state "all cooperate" is 0.48. As a rule, this probability is found to be close to 0.5 at the straight line $\bar{W}=1/2-\bar{\varepsilon}$. Above this line all cooperate, below this line all defect, except the vicinity of the line (of width about 0.1 for $N=10^3$), where the probability of the state "all cooperate" changes continuously from 0 to 1. This means that a manipulation of the initial state $(\bar{\varepsilon},\bar{W})$ modifies the final result, which still remains homogeneous. In particular, when the initial value of $\bar\varepsilon$ is shifted by 0.1 downwards, all defect; upwards, all cooperate. The properties of the boundary $\bar{W}=1/2-\bar{\varepsilon}$ are a consequence of the adopted form of $P(i,j)$, but the homogeneity of stationary states comes from the system dynamics. The results are obtained for $x=0.1$; an increase of $x$ just speeds the process up. For $x=1/2$, the time dependent mean square root of reputation and altruism remain equal, when decreasing to zero.\\

For the option B, we observe the same homogeneous states "all cooperate" or "all defect", but the probabilities of these states are different. Again we use the same initial reputation for all agents, and the same statistics. For each system of $N=10^3$ agents we made three runs, each of 100 timesteps, with different initial conditions: {\it i)} $V_1=-0.5$ and $V_2=0.3$, {\it ii)} $V_1=-0.4$ and $V_2=0.4$, {\it iii)} $V_1=-0.3$ and $V_2=0.5$. While the initial conditions {\it ii)} are neutral, the case {\it i)} promotes defection and the case {\it iii)} promotes cooperation. For the option A, the same initial conditions served to demonstrate the symmetry of two strategies. However, for the option B we observe that in the case {\it ii)} cooperation prevails. The obtained numbers for {\it i)} are $W_i=0$ for all agents, and the average $\varepsilon_i$ for each system is $-.19\pm 0.15$. In the case {\it iii)} all cooperate, and their altruism is +0.5. In the case {\it ii)} we get again two homogeneous states: "all cooperate" with probability 0.88, "all defect" with probability 0.12. In the former state, the altruism of all agents is maximal: 0.5. In the latter, for each system the average $\varepsilon$ is $-0.15\pm.11$. Note that in the option B, the altruism of uncooperative agents remains unchanged, hence its spread in the uncooperative phase. An exemplary plot of the altruism in this case is shown in Fig. 2. Again, these results are obtained for $x=0.1$. When $x$ increases to 1, the average altruism $\varepsilon$ drops to its minimal value -0.5 almost linearly with $x$.\\

For the option C, calculations are made for the three above given initial conditions and the same statistics. In this option, the altruism of each agent remain unchanged, then for theinitial conditions $\it i)$ and $\it iii)$ a permanent bias is present in the system towards defection and cooperation, respectively. However, the system evolution (Eqns. 6 and 7) produces another bias, always towards cooperation. Clearly, there is no homogeneous phase here. The obtained values of mean reputations for each of $10^3$ systems are practically the same. These results are shown in Fig. 3, as dependent on the parameter $z$. To demonstrate the character of the latter bias, we show also in Fig. 4 an example of the plots of probability of a common cooperation (R), of a common defection (U), of cooperating but being defected (S) and of defecting a cooperating co-player (T). Plots of the same character were shown in \cite{gaw} for the symmetric case where altruism is constant. As we see in the plots, in the option C the cooperation is promoted.\\

\section{Discussion}

The result of our simulation is that once the altruism is allowed to evolve, in long time limit the simulated players adopt one strategy, the same for the whole population. This strategy is either to cooperate, or to defect. For the adopted initial distributions of $\varepsilon_i$ and $W_i$, basically the final outcome is determined by the initial mean values $\bar{W} \equiv (V_1+V_2)/2$ and $\bar{\varepsilon} \equiv (E_1+E_2)/2$ as follows: once $\bar{W}+\bar{\varepsilon}>1/2$, the final strategy is to cooperate, otherwise the strategy is to defect. This is true except the case when $\bar{W}+\bar{\varepsilon}\approx 1/2$. In this case it is possible that the whole population defects or cooperates; the respective probabilities vary with $\bar{W}+\bar{\varepsilon}$. This result is new and completely different from the case $x=0$, considered previously \cite{gaw}. It is different also from the results obtained above for the model of indirect reciprocity, where altruism does not vary. In the latter model the mechanism which stabilizes cooperation is that agents with small (bad) reputation, even if defected, do not influence the reputation of co-players. \\

As remarked above, the time evolution of human general attitudes to cooperate is expected to vary slower than the opinions on particular co-players, and it seems reasonable to believe that the former is driven by the latter. We would like to stress that as a rule, what is observed in social phenomena is an interplay of transient effect with different characteristic times. Then, conclusions of modeling should be related rather to the direction of the process than to the stationary state in the long time limit. In particular, our model takes into account a coupling between an agent's experience on the behaviour of the others and the overall willingness of this agent to cooperate. Our results indicate that the feedback is positive; more cooperation bears more altruism what in turn leads to more cooperation. As a rule, an agent's experience that cooperation is met in most cases leads to a general belief that to cooperate is an accepted social norm. Then, in our theories on experimental data we should consider rather the direction of the process than its stationary stage. One of experiments of this kind was conducted in the Swiss army \cite{swiss}, within the Prisoner's Dilemma scheme. There, platoons of males were formed in a random way for 4-week period of officer training. Having finished the training, individuals believed that members of their own platoons were more willing to cooperate, than others. More data on social experiments can be found in \cite{cam}.\\

As we noted in the Introduction, there is no direct one-to-one correspondence between social effects and theories; the same effect can be discussed within more than one theory. It seems worthwhile to refer here again to the Scheff theory of shame and pride \cite{sch,rev}. This theory describes the self-stabilizing consequences of social interaction to a loop of pride and a loop of shame. Again, as in the definition of social norm by Bicchieri \cite{cb}, the crucial factor of interaction is the content of mutual expectation. On the laboratory side, we have the same ambiguity. A recent list of reputation-based experiments was composed by Binglin Gong and Chun-Lei Yang in \cite{fi}. In the same paper a new experiment of this kind is described, and the interpretation provided is again twofold: either the indirect reciprocity, or some "sense of justice" of the participants. Concluding, at the present stage of social theory we have to accept, that both in the case of experiment and of simulation an interpretation is not unique. It seems to us that identifying connections between different effects and/or properties of social systems is less ambiguous.\\

\bigskip

{\bf Acknowledgements.} One of the authors (K.K.) is grateful to Hisashi Ohtsuki for a heplful discussion. The research is partially supported within the FP7 project SOCIONICAL, No. 231288.

\end{document}